\documentclass[a4paper]{spie}  %>>> use this instead for A4 paper
\usepackage[]{graphicx}

\title{XIAO:  a soft X-ray telescope
for the SVOM Mission}

\author{S. Mereghetti\supit{a},
 A. De Luca\supit{a},
 M. Fiorini\supit{a},
 N. La Palombara\supit{a},
 A. Tiengo\supit{a},
 M. Uslenghi\supit{a},
 G. Vianello\supit{a},
 G. Villa\supit{a},
 A. Zambra\supit{a},
 A. Moretti\supit{b},
 S. Basso\supit{b},
 S. Campana\supit{b},
 P. Conconi\supit{b},
 V. Cotroneo\supit{b},
 S. Covino\supit{b},
 G. Ghirlanda\supit{b},
 G. Ghisellini\supit{b},
 G. Pareschi\supit{b},
 G. Tagliaferri\supit{b},
 M. Trifoglio\supit{c},
 L. Amati\supit{c},
 F. Gianotti\supit{c},
 L. Stella\supit{d},
 A. Antonelli\supit{d},
 B. Cordier\supit{e},
 D. G\"{o}tz\supit{e}
 \skiplinehalf
\supit{a}INAF - IASF Milano, v. E. Bassini 15, 20133 Milano, Italy; \\
\supit{b}INAF - Osservatorio Astronomico di Brera, via Bianchi 46,
23807 Merate , Italy; \\
\supit{c}INAF - IASF Bologna, v. Gobetti 101, 40129 Bologna, Italy; \\
\supit{d}INAF - Osservatorio Astronomico di Roma, via Frascati 33,
00040 Monteporzio Catone,  Italy;\\
\supit{e} CEA Saclay, Service d'Astrophysique, F-91191
Gif-sur-Yvette, France.}
\authorinfo{Further author information: (Send correspondence to S.Mereghetti)\\
E-mail: sandro@iasf-milano.inaf.it, Telephone: +39 02 23699323}
\def\ltsima{$\; \buildrel < \over \sim \;$}
\def\lsim{\lower.5ex\hbox{\ltsima}}
\def\gtsima{$\; \buildrel > \over \sim \;$}
\def\gsim{\lower.5ex\hbox{\gtsima}}

\def\aj {AJ}
\def\mnras {MNRAS}
\def\aap {A\&A}
\def\aapr {A\&AR}
\def\apj {ApJ}
\def\apjl {ApJL}
\def\nat {Nature}

\begin{document}
  \maketitle

\textbf{Presented at: ``SPIE Astronomical Telescopes and
Instrumentation 2008'' - Marseille, June 23-28, 2008}

%%%%%%%%%%%%%%%%%%%%%%%%%%%%%%%%%%%%%%%%%%%%%%%%%%%%%%%%%%%%%
\begin{abstract}
The study of Gamma-ray bursts (GRBs) is a key field to expand our
understanding of several astrophysical and cosmological phenomena.
SVOM is a Chinese-French Mission which will permit to detect and
rapidly locate   GRBs, in particular those at high redshift, and
to study their multiwavelength emission. The SVOM satellite, to be
launched in 2013,  will carry wide field instruments operating in
the X-/$\gamma$-ray band and narrow field optical and soft X-ray
telescopes. Here we describe a small soft X-ray telescope (XIAO)
proposed as an Italian contribution to the SVOM mission. Thanks to
a grazing incidence X--ray telescope with effective area of
$\sim$120 cm$^{2}$ and a short focal length, coupled to a very
compact, low noise, fast read out CCD camera, XIAO  can
substantially contribute to the overall SVOM capabilities for both
GRB and non-GRB science.

\end{abstract}

%>>>> Include a list of keywords after the abstract


%%%%%%%%%%%%%%%%%%%%%%%%%%%%%%%%%%%%%%%%%%%%%%%%%%%%%%%%%%%%%
%\section{INTRODUCTION}
%\label{sec:intro}  % \label{} allows reference to this section

\section{The SVOM Mission}

The study of Gamma-ray bursts (GRBs), extremely luminous transient
sources appearing in the sky when black holes are born in the
explosions of massive stars or in the merging of compact stellar
objects, has become a key field to expand our understanding of
several astrophysical and cosmological phenomena. These  include,
e.g., the evolution of the  young universe, the history of star
formation, the metal enrichment of galaxies, the mechanisms
driving supernova explosions, the physics of ultra-relativistic
shocks.

SVOM (Space-based multi-band astronomical Variable Objects
Monitor) is a mission, developed in collaboration by the Chinese
and French space agencies (CNSA and CNES), for the study of GRBs
and other high-energy transients. SVOM is designed  to detect and
rapidly locate all kinds of GRBs, in particular those at high
redshift, and to study their emission on a broad spectral range,
from the visible to the MeV region. This will be possible thanks
to a satellite payload composed of  wide field X-/$\gamma$-ray
instruments and narrow field optical and soft X-ray telescopes,
complemented by dedicated ground telescopes and a system for rapid
distribution of the GRB positions.

Building on the successful experience of Swift\cite{geh04}, the
SVOM  operations are based on the following steps: \textsl{(i)}
GRB detection with a wide field gamma-ray imaging instrument able
to derive on-board its localization with a few arcmin
precision\cite{sch07} ; \textsl{(ii)} the GRB position is
immediately transmitted to ground through a network of ground
stations and at the same time the satellite slews rapidly to
position the GRB in the narrow fields of view of its  X-ray and
optical telescopes, which will study the afterglow and provide
refined coordinates.

SVOM will adopt an optimized observation  strategy \cite{cor08},
based on antisolar pointing and avoidance of the galactic plane,
in order to permit follow-up observations with large telescopes
and maximize the number of redshift measurements. The knowledge of
redshift and the determination of the spectral shape over  an
extended energy range are in fact essential to derive the bursts
energetics and to study the empirical correlations used for
cosmological studies.

The SVOM satellite will carry the following instruments:

\begin{itemize}

\item \textit{Camera X and Gamma (CXG)}: a wide field instrument
operating in the 4-300 keV energy range, with a field of view of 2
sr and a location accuracy of several arcmin.  This instrument,
providing the GRB triggers and initial localizations, is based on
an array of CdTe pixels with a sensitive area of 1000 cm$^{2}$
coupled to a coded mask aperture. It is based on the instrument
originally proposed for the Eclairs satellite \cite{sch06}.

\item \textit{XIAO}: a narrow field soft X--ray telescope to
locate and study the GRB afterglows (see next Section).

\item  \textit{Visual Telescope (VT)}: an optical telescope
operating in the 400-950 nm range, with a field of view of
$\sim$21$'$, reaching a sensitivity of V$\sim$23 magnitudes in 300
s exposure times.

\item \textit{Gamma-ray Burst Monitor (GRM)}: a non-imaging
spectrometer to measure the GRB spectra in the 50 keV-100 MeV
energy range over a wide field of view

\end{itemize}

SVOM will be placed in a near earth orbit, of $\sim$600 km
altitude and \ltsima30$^{\circ}$   inclination. The  GRB
coordinates, and a small set of relevant information, will be
transmitted on ground in real time by means of a network of VHF
stations, as successfully done in the past by the HETE-2
satellite. The bulk of the data will be downloaded (a few times
per day) when the satellite is in contact with the main ground
station(s). The SVOM satellite will be launched in 2013, with the
goal to  detect and precisely locate about 200 GRBs in a nominal
mission duration  of 2.5 years.

%%%%%%%%%%%%%%%%%%%%%%%%%%%%%%%%%%%%%%%%%%%%%%%%%%%%%%%%%%%%%
\section{XIAO scientific performances}

%%-----------------------------------------------------------
%\subsection{}
\label{sec:title}

XIAO (X-ray Imager for Afterglow Observations) is a small and
light X-ray telescope designed with the main objective of
significantly improving the GRB locations obtained on board the
satellite, through a prompt identification of the afterglows. This
can be achieved with a grazing incidence mirror operating in the
soft X--ray band.

\begin{figure}
   \begin{center}
   \begin{tabular}{c}
   \includegraphics[height=7cm]{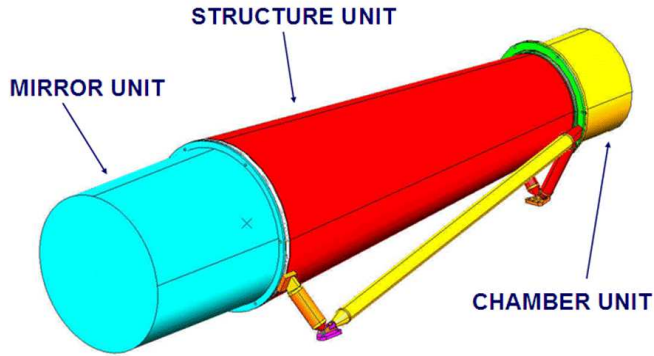}
   \end{tabular}
   \end{center}
   \caption[example]
%>>>> use \label inside caption to get Fig. number with \ref{}
   { \label{fig:xiao}
Conceptual design showing the main elements of the XIAO telescope.
The radiator and the digital electronic box are not shown.
(courtesy A. Stevoli - TAS-I) }
   \end{figure}

Taking into account the tight constraints of mass and dimensions,
an optimized design based on a Wolter I  mirror with a short focal
length  coupled to a fast read out, low noise CCD detector has
been chosen. A light structure in carbon fiber will connect the
Mirror Module and CCD Camera Units, providing the required
stiffness and shielding the CCD from optical radiation. The
structure will also provide the interfaces to mount the XIAO
telescope on the satellite Payload  Interface Module.  A dedicated
electronics will compute  in real time the source positions to be
immediately transmitted to the ground. A conceptual design of the
main XIAO telescope elements is shown in Fig.~\ref{fig:xiao}.

%-------------
%-------------

XIAO is required to cover a field of view of $\sim$25 arcmin
diameter, in order to include, with a safe margin, the initial GRB
error region provided by the CXG  instrument. A moderate angular
resolution is adequate, considering that most of the observed
bursts will lie at high galactic latitude, where no source
confusion is expected. This translates in a requirement of
$\sim30''$  (Half Energy Diameter, HED) for the XIAO point spread
function.

\begin{figure}
   \begin{center}
   \begin{tabular}{c}
   \includegraphics[height=7cm]{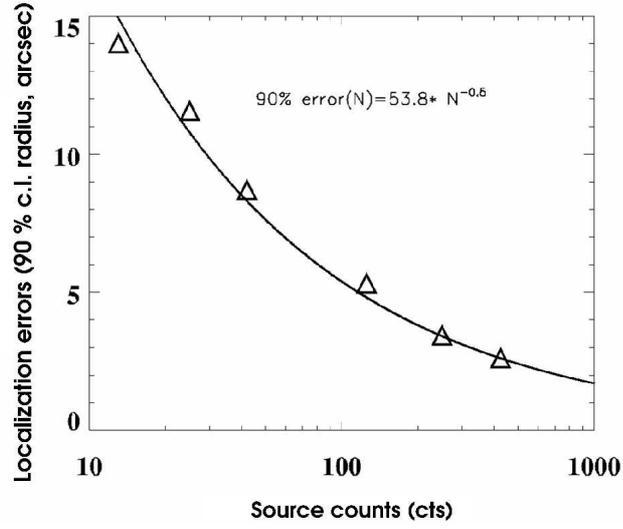}
   \end{tabular}
   \end{center}
   \caption[example]
    { \label{fig:poserr}
XIAO source location accuracy as a function of the number of
detected photons. The data points have been obtained with
simulations assuming a point spread function with HED=30$''$. The
line is the best fit with a function $\sigma_{POS} \propto
N^{-0.5}$. }
   \end{figure}

On the other hand, a good localization accuracy is one of the main
drivers of the XIAO design. The source localization accuracy,
$\sigma_{POS}$, is linked to the width of the instrument point
spread function by  $\sigma_{POS}\sim$
$k\times\frac{\sigma_{HED}}{\sqrt{N}}$,  where $k$ is a constant
depending on the instrument point spread function, $\sigma_{HED}$
is the width of the point spread function,  and $N$ is the number
of detected photons. This relation, supported by experience with
similar X--ray telescopes, is confirmed by simulations of the XIAO
instrument, as shown in Fig.~\ref{fig:poserr}. If only statistical
errors are considered, localizations at the arcsecond level can be
obtained with XIAO as soon as few hundreds of  X--ray afterglow
photons are collected. In practice, for most cases the
localization accuracy achievable on board will be limited by
systematics affecting the attitude reconstruction. Quick look
ground analysis, also exploiting the VT data, will permit a
reduction of the systematic effects and lead to more accurate
positions.

\begin{figure}
   \begin{center}
   \begin{tabular}{c}
    \includegraphics[width=7cm , angle=90]{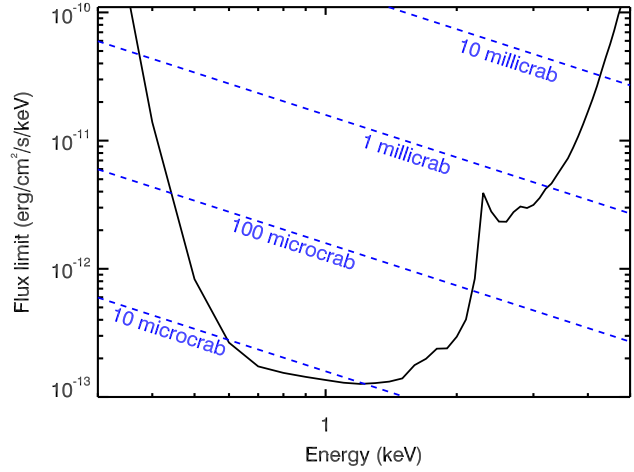}
   \end{tabular}
   \end{center}
   \caption[example]{ \label{fig:sensi}
Expected sensitivity of the XIAO telescope (5$\sigma$ detection in
10 ks), computed assuming a 30$''$ Half Energy Diameter,
% a background rate of 2 10$^{-3}$ counts cm$^{-2}$ s$^{-1}$ keV$^{-1}$,
a source extraction circle of 15$''$ radius,  and an
energy bin $\Delta$E=E/2.}
   \end{figure}

The expected throughput of XIAO is a function of the source
spectral properties, and in particular of the interstellar
absorption. Assuming a power law spectrum with photon index
$\Gamma$=2 and N$_{H}$ = 10$^{21}$ cm$^{-2}$, with the mirror
effective area and the CCD Camera described below, we expect 1
count s$^{-1}$ for a 2-10 keV flux of 3 10$^{-11}$ erg cm$^{-2}$
s$^{-1}$. The expected sensitivity is shown in
Fig.~\ref{fig:sensi}. Note that, thanks to the short focal length,
the particle induced background in the resulting very small source
extraction region,  will be very small. Thus, even for relatively
long exposure times, the XIAO sensitivity will not be
background-dominated.

\begin{figure}
   \begin{center}
   \begin{tabular}{c}
  \includegraphics[width=11cm]{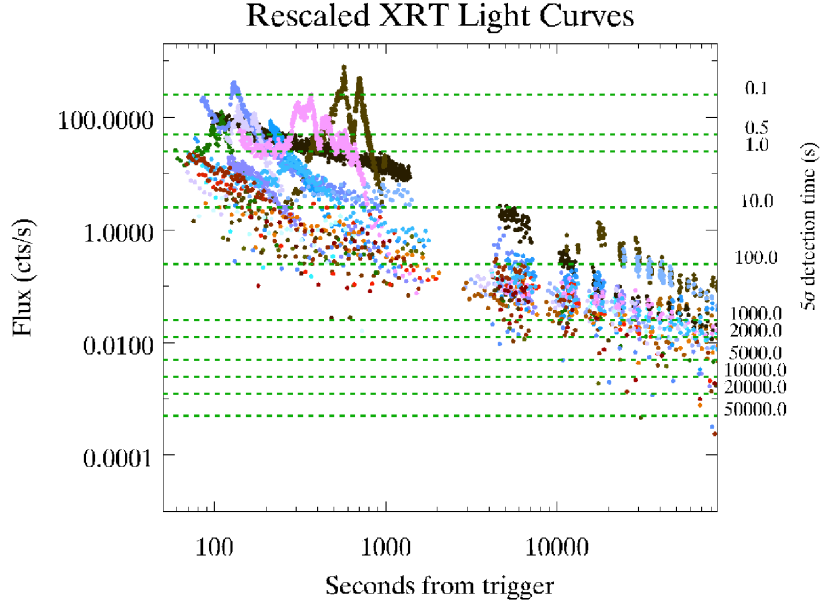} \\
   \end{tabular}
   \end{center}
   \caption[example]{ \label{fig:aft}
Expected XIAO light curves for a sample of representative GRB
afterglows. The horizontal lines give the XIAO 5$\sigma$
sensitivities corresponding to the integration times marked on the
right axis.}
   \end{figure}

In Fig.~\ref{fig:aft} we have plotted  a  sample of X-ray
afterglow light curves converting the fluxes observed with the
Swift/XRT instrument to the expected XIAO count rate. In this
conversion we have properly taken into account the  spectral
parameters of each afterglow. The figure shows that XIAO has a
sensitivity adequate to  provide precise localizations for most
GRBs, considering that X-ray afterglows are observed in $\sim$95\%
of the GRBs. Detailed studies of the brightest afterglows will
also be possible during follow-up observations.

Table \ref{tab:xiao} summarizes the scientific performances
expected for XIAO.

%-------------
%-------------

%-------------
%-------------

%-------------
%-------------

%% This table is carefully placed in the source file to make
%% it appear at bottom of page, but above the footnotes.
%% Use of [h] in following command forces table to appear "here".
\begin{table}[h]
\caption{Expected performances of the XIAO soft X-ray telescope}
\label{tab:xiao}
\begin{center}
\begin{tabular}{|l|l|} %% this creates two columns
%% |l|l| to left justify each column entry
%% |c|c| to center each column entry
%% use of \rule[]{}{} below opens up each row
\hline
\rule[-1ex]{0pt}{3.5ex}  Energy Range & 0.5 -- 2 keV   \\
\hline
\rule[-1ex]{0pt}{3.5ex}  Field of view  &  27$'$ diameter  \\
\hline
\rule[-1ex]{0pt}{3.5ex}  Angular resolution &  30$''$ (Half Energy Diameter)  \\
\hline
\rule[-1ex]{0pt}{3.5ex}  Location accuracy  & $\sim$10$''$ for a source at 5$\sigma$  \\
\hline
\rule[-1ex]{0pt}{3.5ex}              &   \lsim5$''$ for a source at $>$10$\sigma$ \\
\hline
\rule[-1ex]{0pt}{3.5ex}  Sensitivity (5$\sigma$)  &   $\sim$10 mCrab in 10 s  \\
\hline
\rule[-1ex]{0pt}{3.5ex}        &  $\sim$10-20 $\mu$Crab in 10 ks    \\
\hline
\rule[-1ex]{0pt}{3.5ex}  Energy resolution  & $\sim$150 eV (FWHM at 1.5 keV) \\
\hline
\rule[-1ex]{0pt}{3.5ex}  Time resolution  &  $\sim$10 ms \\
\hline
\end{tabular}
\end{center}
\end{table}

%%  Use following command to specify that graphics file is in
%%  a directory other than this LaTeX source file.
%%  Note use of / to separate subdirectories, for UNIX and Windows OS.
%%\graphicspath{{H:/HANSON/SPIESTY/}}
%% tabular environment useful for creating an array of images

%%%%%%%%%%%%%%%%%%%%%%%%%%%%%%%%%%%%%%%%%%%%%%%%%%%%%%%%%%%%%
\section{XIAO baseline design}

%%-----------------------------------------------------------
\subsection{Mirror Module Unit}
\label{sec:title}

The XIAO Mirror Module Unit comprises the X--ray optics with the
associated mechanical structure, a thermal baffle, and a front
cover.

The optical design of the mirrors consists of 6 Wolter I shells
with focal length of 830 mm.  The 6 shells have the same length
equally shared by parabola (300 mm) and hyperbole (300 mm).
Diameters range from 250 to 144 mm, and the thickness is $\sim$300
$\mu$m for all the shells.  The present design is the result of an
optimization procedure where the most severe requirements were the
total weight allocated to the shells (\lsim5 kg) and the maximum
length of the telescope. %($<$ 120 cm).
The optimization has been performed by evaluating the telescope
effective area over the field of view weighted by the GRB position
accuracy expected from the CXG (approximately a Gaussian with
$\sigma\sim$4$'$). The short focal length gives a plate scale
$\sim$4$'$ mm$^{-1}$, thus the required field of view can be
efficiently covered with a very small detector.

%-------------
   \begin{figure}
   \begin{center}
   \begin{tabular}{c}
   \includegraphics[width=12cm]{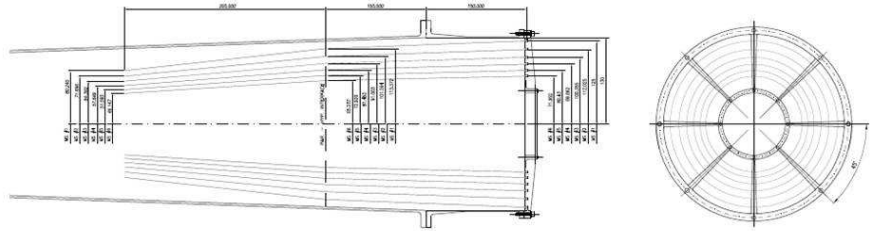} \\
   \end{tabular}
   \end{center}
   \caption[example]
    { \label{fig:model} Preliminary design of the
    XIAO X--ray optics (courtesy R.Buzzi - BCV). }
   \end{figure}
%-------------

\begin{figure}
   \begin{center}
   \begin{tabular}{c}
   \includegraphics[height=7cm]{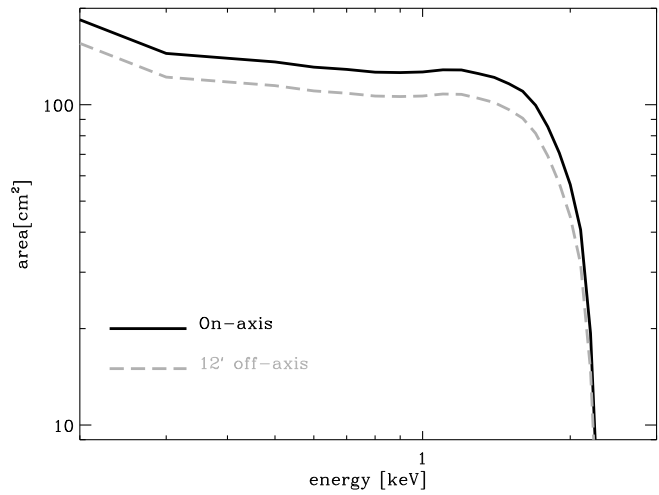}
   \end{tabular}
   \end{center}
   \caption[example]
    { \label{fig:area} The effective area expected for the current optical
design, taking into account a loss of 10\% due to the spider
occultation. The solid line is for on-axis sources, while the
dashed line refers to an off-axis angle of 12$'$. }
   \end{figure}

The mirrors will be produced by replicating superpolished mandrels
by means of electroformed Nickel shells. This is a well
consolidated technology developed for BeppoSAX and subsequently
applied with success to XMM-Newton and Swift/XRT. The shells will
be integrated by means of a single spider with eight arms
positioned on the front pupil as shown in Fig.~\ref{fig:model}.
With a shell coating in gold, this configuration is providing the
effective area shown in Fig.~\ref{fig:area}.

A thermal baffle is placed in front of the mirror to keep it at an
operating temperature of $\sim$20$^{\circ}$ avoiding gradients
that would lead to deformation of the optics and a degradation of
the imaging performance. A   cover is required to avoid
contamination during ground activities, launch  and early orbital
phases.

%-------------
%-------------

%%-----------------------------------------------------------
\subsection{CCD Camera Unit}
\label{sec:title}

The X--ray photons focused by the mirror unit will be detected by
a small, fast scan, X--ray CCD operating in photon counting mode
and placed in the focal plane. A CCD Camera Unit (CCU) will host
the detector and all the elements required for operating the
sensor. The CCU will include: a detector sub-system (CCD and
thermal control based on a Thermo Electric Cooler (TEC)), the
Proximity and Front-End Electronics, and a shielding sub-system
(proton shield and shutter).

The baseline detector has been preliminarily  identified in a
customized, X--ray optimized, version of the e2V CCD67 sensor. The
device is a frame transfer CCD with active area of 256$\times$256
pixels (pixel size 26 $\mu$m, corresponding to 6.5$''$ for the
XIAO focal length). This device is characterized by being suitable
for high frame rate readout, thanks  to the two  output nodes and
amplifiers which allow readout frequency up to 5 MHz
(corresponding to 150 frames s$^{-1}$). The sensor is available in
a back-illuminated, thinned version. Due to the soft energy range
of the telescope (\lsim2 keV), a thickness of $\sim$25 $\mu$m has
been chosen.

In order to allow photon counting operation with low pile-up even
in the early stages of the GRB afterglow, when the incoming flux
can be high, the CCD should be read at the highest frame rate.
However, a trade-off has to be made with the readout noise: higher
noise affects the performance of the detector not only in terms of
energy resolution, which is a secondary parameter for XIAO, but,
most important, in terms of sensitivity at low energy. Based on
these considerations, a frame rate of 100 frames s$^{-1}$ is
considered the goal. Taking into account the PSF of the mirrors,
this translates in a 10\% loss of linearity under a count rate of
$^{\sim}$200 counts s$^{-1}$ from a point-like source.

Active cooling to an operative value of about --65$^{\circ}$ will
be provided by means of a TEC coupled through a cold finger to a
radiator mounted on the cold side of the satellite.  A fixed
proton shield will protect the detector from the flux of charged
particles, and a fixed optical/UV blocking filter will be placed
above the CCD, in order to reduce the optical loading of the
detector. Moreover, a shutter will allow closing the camera,
preventing the electromagnetic radiation to reach the detector
(for testing and calibration purposes, other than for safety). The
shutter will also provide protection from low energy particles
during SAA crossings. A calibration source could be allocated on
the backside of the shutter seen by the CCD.

The camera will be enclosed in a vacuum housing, to allow ground
tests and launch in vacuum. The housing will be hermetically
sealed by a door, which will be operated on-ground during tests
and ``one shot'' on-orbit.  A vacuum valve will allow on ground
evacuation or gas filling, e.g. with dry nitrogen during storage.
The Camera Unit will also contain part of the electronics, as
described below.

%%-----------------------------------------------------------
\subsection{Electronics}
\label{sec:title}

The XIAO electronics subsystem includes the front end electronics
of the CCD and a Digital Processing Unit (DPU), which controls the
XIAO telescope, handling instrument power distribution, telemetry
and telecommand management, scientific data acquisition and
processing, and I/F management.

\subsubsection{Front End Electronics} \label{sec:FEE}

The CCD Front End Electronics (FEE) will include two main blocks:
an Analog Front-End Electronics (AFEE) and a  Digital Front-End
Electronics (DFEE).

The AFEE will include the CCD bias generator, clock drivers, and
two analog signal processing chains (one for each output node).
The main guideline in the AFEE design is to maintain its
contribution to the readout noise negligible respect to the CCD
on-chip amplifier noise. This translates in the request of noise
\lsim10 electrons rms.  Part of the electronics will be located
directly near the CCD to limit the noise, in particular:
preamplifiers, bias generators and clock drivers will be located
as close as possible to the CCD (Proximity Electronics).
Electrical connections between the proximity FEE board and the
sensor will be made via two flexi connectors that also provide a
thermal break.

The DFEE will include the CCD controller and the components for
the real-time pre-processing of the images. The latter will work
only when the camera is operated in photon counting mode and will
implement the following tasks: \textit{(i)} ``valid" X--ray events
pattern  recognition
   (rejecting cosmic ray traces and other  contaminants);
\textit{(ii)} bright pixel rejection; \textit{(iii)} events
coordinates computation; \textit{(iv)} energy   evaluation (in
case of splitted events by summing energy deposited in all the
pixels involved). The output of the image pre--processing is a
list of X--ray events coordinates, with additional information
about timing and energy, which is sent to the DPU.

\subsubsection{Data Processing Unit}
\label{sec:FEE}

The DPU will provide all the usual services required to operate
the instrument, such as management of the secondary voltage lines,
actuators (TEC, heaters, sensors, valves and mechanisms),
management of data and housekeeping, etc... A microprocessor based
block will be in charge of TM/TC management (from/to S/C and CCU)
and thermal control.

In addition to these ``standard'' functions, the DPU is in charge
of the real time computation of the GRB coordinates when XIAO is
in the ``GRB Localization" mode.  This is done be elaborating the
event coordinates as generated by the DFEE (which will operate in
``photon counting mode") in order to localize the spot
corresponding to the GRB afterglow and then computing the
coordinates (and associated uncertainties) of its centroid.

\section{Science with XIAO}

\subsection{Scientific objectives for GRBs}

To reach the main SVOM  scientific goals (e.g. population studies
and cosmology with GRBs), the presence on board of an X-ray
telescope like XIAO is mandatory. In fact an X-ray telescope
observing the afterglows is needed to locate GRBs with sufficient
accuracy (down to few arcsec) required to promptly identify an
optical counterpart and measure the redshift. XIAO will provide an
intermediate step between the first localizations at several
arcmin level given by the CXG and the precise localizations that
can in principle be achieved with the optical telescope. It must
also be considered that about half of the bursts are optically
dark \cite{nar08}. Particularly interesting will be the z$>$5
bursts (drop-outs in optical) that SVOM is expected to detect
thanks to the CXG energy range extending to low energy ($\sim$4
keV). In addition to the GRB localization tasks,  there are also
several specific objectives for which XIAO is essential. In the
following we highlight a few of them.

XIAO extends the SVOM spectral coverage down to at least 0.5 keV,
thus complementing the energy range of the CXG and GRM. For very
long bursts, or bursts triggered on precursors, XIAO will see the
prompt emission. This will allow to constrain  the spectral
parameters including the peak energy \cite{ama02} which is the
crucial parameter to understand the GRB physics and do cosmology.
The study of the  GRB class of X--ray Flashes \cite{sak08}  will
largely benefit from XIAO observations.

Spectral fitting of the X-ray afterglow with XIAO will determine
the GRB intrinsic hydrogen column. Swift showed that this can be a
tool to identify high redshift bursts \cite{gru07}  since the
photons from these sources are emitted at higher energies, less
affected by absorption. The pre-selection of high redshift
candidates can be useful to tune the ground-based and space follow
up.

XIAO will shed light on the physics of the afterglow which is now
an open issue. Indeed, several bursts show a different behavior in
the optical and X-rays,  still not convincingly explained
\cite{ghi07}. Combined with the VT and ground-based optical/IR
follow up, XIAO will give light curves and time resolved spectra
of the different (early and late) phases of the afterglow
emission. In particular, it will be possible to study the issue of
(a)chromatic breaks, crucial to derive the collimation angle of
the jet (hence its true energy)\cite{ghir07}.

XIAO will detect the X-ray flares superimposed to the continuous
afterglow emission. Their nature is not understood: they could
correspond to the late time accretion of a fragmented disk or to
slow shells \cite{obr06}. In either case the energetics and
spectra of the flares can distinguish among the proposed
interpretations. Complemented by the high-energy detection by the
CXG and GRM, XIAO will prove if flares have the same nature of
earlier prompt pulses, thus unveiling the possible accretion
modes/regimes at the hearth of the central engine \cite{laz08}.

A few cases of low luminosity very long GRBs have been discovered
recently and have been proposed to be a distinct class of bursts,
possibly with a different central engine: a magnetar instead of a
black hole \cite{tom07}. They emit most of their energy in the
soft X-ray band  where XIAO is  sensitive. One of these bursts
(GRB 060218 \cite{cam06}) also showed a thermal black body
component with a temperature of 0.2 keV whose nature is debated
(supernova shock breakout or fireball matter-radiation
decoupling).

XIAO will give a series of snapshots of the chemical composition
of the circumburst medium (through absorption edges and features
\cite{cam08}). This is particularly important because it is
related to the apparent contradiction of a uniform circumburst
medium (as suggested by the present observations) and the
expectation of a stratified wind profile (produced by a Wolf-Rayet
progenitor star \cite{che00}). Moreover, it gives information on
the dust-to-gas ratio in the vicinity of the bursts related to the
optical extinction and X-ray absorption.

XIAO is important to detect and characterize low luminosity, long,
soft and nearby GRBs of the kind of GRB 060218 \cite{cam06}.
Thanks to its low energy range, the SVOM CXG can reveal a larger
number of these events, that were not detected by BATSE and only
marginally by BAT/Swift, and XIAO will be crucial in localizing
and studying their X-ray emission. The importance to study this
class of bursts stems out from the recent proposed association
\cite{ghi08} of ultra high energy cosmic rays (E$>$57 EeV). If
true this would make these sources the most energetic accelerators
of cosmic particles, solving a more than 40-years-long debate.

\begin{figure}
   \begin{center}
   \begin{tabular}{c}
   \includegraphics[width=10cm]{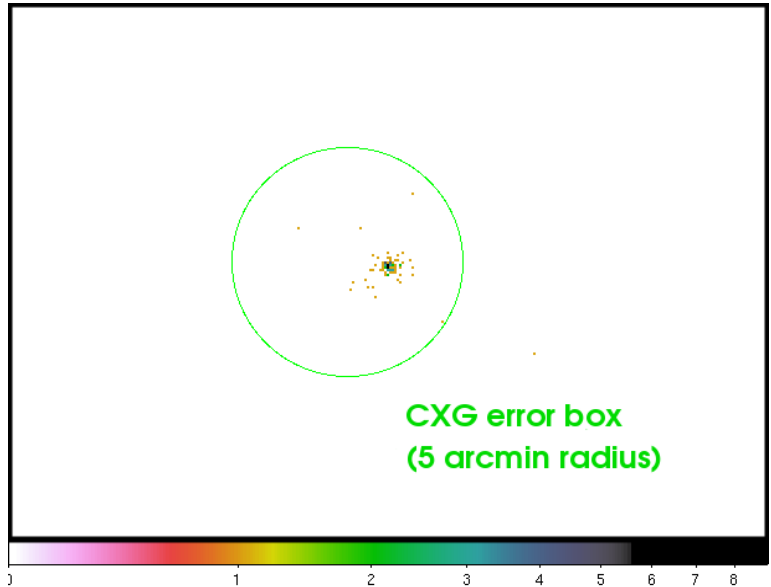} \\
   \end{tabular}
   \end{center}
   \caption[example]
    { \label{fig:grb} Simulated image of the afterglow of GRB 050904 as
would have seen by XIAO $160$ s after the trigger.  The green
circle shows a representative example of the initial localization
accuracy obtained with the CXG (radius 5$'$ in the case of this
high fluence GRB).  With this first image (2.5 s integration time)
XIAO gives an error circle with 3.7$''$ radius.  }
   \end{figure}

Finally, as an example of  the XIAO capabilities,  we present a
simulated observation of the  GRB 050904 \cite{tag05}, the burst
with the highest measured redshift (z=6.3). We used the X-ray
light curve and spectrum measured by Swift/XRT and assumed a slew
time of 160 s to reach a stabilized pointing with the GRB error
region in the XIAO field of view. Integrating the first 2.5
seconds of the XIAO data we obtain the image shown in
Fig.~\ref{fig:grb}, which contains about 120 afterglow photons.
Analysis of this image with a simple centroid algorithm,
representative of what could be implemented on board, results in a
localization uncertainty of $\sim$4$''$ radius.
%This is compared with the CXG error circle of about 5
%arcmin expected for such a high fluence GRB. We note that t
The high read out frequency of the XIAO CCD allows to carry out
all the observations in photon counting mode. Thus, contrary to
the case of Swift/XRT, full timing and energy information is
available for the counts in the first image. The Swift spectral
analysis of this bright afterglow was complicated by the effects
of pile-up and changing CCD modes. These issues will be much less
important for XIAO.

%-------------
%-------------

\subsection{Non-GRB Science}

Based on current estimates for GRB trigger rate and observing
efficiency, we may expect that only a  fraction of the SVOM
observing time will be devoted to GRBs. To maximize the science
outcome of the mission, it is thus very important to plan a
rewarding program of non-GRB science observations, that in any
case, should not impact on the mission requirements for GRB
detections and follow-up study.

Therefore non-GRB observations should (i) comply with the SVOM
optimized pointing strategy and (ii) be interrupted whenever a GRB
trigger occurs. The first point limits the fraction of the sky
which may be investigated with a narrow-field instrument such as
XIAO. On the other hand, it allows to perform observations of the
same target for a  rather long time interval (up to more than one
month). Thus, multiwavelength, long term, almost uninterrupted
monitoring of variable sources can be identified as a unique
capability of SVOM, setting the case for unprecedented studies of
different classes of astrophysical sources.

The coordinated use of XIAO and the VT on board SVOM will allow to
perform truly simultaneous X--ray and optical observations,
something which is not easily implemented by coordinating
satellite and ground based telescopes.  It will thus be  possible
to assess the amplitude of variability over an extended energy
range, as well as possible correlations between the different
bands. The XIAO telescope, thanks to its good sensitivity
capabilities, can drive the selection of the main topics of
non-GRB science for SVOM. We  describe in the following a few of
such possible science themes.

\subsubsection{Active Galactic Nuclei}

Active Galactic Nuclei come in a large diversity of
manifestations. Time variability and complex, rich spectral energy
distributions (SED) are the rule. Among AGN targets for XIAO, we
may list sources as different as blazars, Narrow-Line Seyfert 1
(NLS1) galaxies and Low-Ionization Narrow Emission-line Regions
(LINERs).

Blazars are very powerful sources. Their emission, dominated by an
ultrarelativistic jet powered by the central supermassive black
hole, extends from radio to TeV energies and  shows a dramatic
variability at all frequencies. Several bright blazars will be
easily observed by XIAO. As an example, correlating the sky region
accessible to SVOM in the optimal pointing strategy with sources
in the ROSAT Bright Survey (hereafter RBS\cite{sch00}), we found
16 BL-Lac sources and 10 flat Spectrum Radio Quasars with a
visibility ranging from 6 to 36 days.  XIAO observations, coupled
to the VT data, will probe the SED on a day-by-day basis which, in
synergy with simultaneous available GLAST observations, will be
crucial to test and discriminate models for jet high energy
emission.

NLS1 are poorly understood sources which are believed to host
relatively low-mass black holes, experiencing a very high
accretion rate. Such a view will be tested with XIAO and the VT
thanks to systematic monitoring of dozens of NLS1 - including,
e.g., 7 sources from the RBS.

LINERs are low-luminosity sources, possibly a scaled-down version
of Seyfert Galaxies (as for the accretion rate). It is not even
clear if they are 'Active' at all. XIAO and VT data will assess
the SED and its variability (if any) for at least 10 sources with
a flux larger than a few 10$^{-13}$ erg cm$^{-2}$ s$^{-1}$ (9-21
days of visibility) and will allow to assess their nature.

\subsubsection{Active Stars}

Stellar flares are the most dramatic manifestation of magnetic
activity in stellar coronae \cite{gud04}. The study of stellar
flares is crucial in order to understand the processes driving the
generation of  stellar magnetic fields, stellar magnetic activity,
and the mechanisms of coronal heating. Observations of flares from
stars of different spectral classes allow to investigate the
dependence of such mechanisms on stellar parameters such as mass,
temperature, gravity, rotation period, age.

SVOM will allow to collect simultaneous high temporal resolution
data in the optical and soft X-ray range for a significant sample
of stellar flares. Up to now, observations of this kind have been
possible only in very few cases,  owing to the  difficulties in
coordinating strictly simultaneous observations with satellite and
ground telescopes. The study of different time scales for the
flares as a function of energy, as well as of the broad band
time-resolved spectral shape, will be very important to test the
current models. SVOM will also be able to detect superflares and
to perform multiwavelength follow-up studies of such rare, very
energetic events, which show non-thermal emission up to $\sim$200
keV. There are 50 bright stars in the RBS with visibility in the
7-29 day range, according to the SVOM nominal pointing law. It
will be particularly interesting to monitor the most active
members of such sample. For instance, dMe stars and RS CVn
systems, with a rate of detectable flares up to $\sim$1 per hour,
will be excellent targets for XIAO.

\subsubsection{Cataclysmic variables}

Cataclysmic variables (CVs) are binary systems featuring a white
dwarf accreting matter from a companion star. There is a rich
phenomenological variety of CVs, mainly depending on the intensity
of the magnetic field of the accreting white dwarf. In any case,
CVs emit from the infrared to hard X-rays with a dramatic time
variability \cite{kuu06}. CVs are valuable laboratories to study
accretion mechanisms in a large range of physical conditions (as
for the magnetic fields and the accretion rates). Within the sky
region accessible with the optimized antisolar pointing, there are
42 CVs listed in the CV catalog \cite{dow06}. Six of them are very
bright in X-rays and are listed in the RASS-BSC. Among CVs, dwarf
novae will be an interesting target. Such sources show  outbursts
with a recurrence time scale ranging from about one week to
several months. The standard model - elaborated on the basis of
optical data alone - explains the large outbursts with the
development and propagation of an instability in the accretion
disc. SVOM will allow simultaneous observations from the IR to
hard X-rays, probing essentially all the emitting regions of such
systems (from the outer region of the disk to the so-called
boundary layer). In particular, the observation of the onset of
the outburst will be crucial to test and improve the current disk
instability model.

%%%%%%%%%%%%%%%%%%%%%%%%%%%%%%%%%%%%%%%%%%%%%%%%%%%%
%\appendix    %>>>> this command starts appendixes
%%%%%%%%%%%%%%%%%%%%%%%%%%%%%%%%%%%%%%%%%%%%%%%%%%%%

%%%%%%%%%%%%%%%%%%%%%%%%%%%%%%%%%%%%%%%%%%%%%%%%%%%%%%%%%%%%%
\acknowledgments     %>>>> equivalent to \section*{ACKNOWLEDGMENTS}

This work has been partially supported by ASI.

%%%%%%%%%%%%%%%%%%%%%%%%%%%%%%%%%%%%%%%%%%%%%%%%%%%%%%%%%%%%%
%%%%% References %%%%%

%\bibliography{XIAO}   %>>>> bibliography data in report.bib

\begin{thebibliography}{10}

\bibitem{geh04}
{Gehrels}, N., {Chincarini}, G., {Giommi}, P., {Mason}, K.~O.,
{Nousek}, J.~A.,
  {Wells}, A.~A., {White}, N.~E., {Barthelmy}, S.~D., {Burrows}, D.~N.,
  {Cominsky}, L.~R., {Hurley}, K.~C., {Marshall}, F.~E., {M{\'e}sz{\'a}ros},
  P., {Roming}, P.~W.~A., {Angelini}, L., {Barbier}, L.~M., {Belloni}, T.,
  {Campana}, S., {Caraveo}, P.~A., {Chester}, M.~M., {Citterio}, O., {Cline},
  T.~L., {Cropper}, M.~S., {Cummings}, J.~R., {Dean}, A.~J., {Feigelson},
  E.~D., {Fenimore}, E.~E., {Frail}, D.~A., {Fruchter}, A.~S., {Garmire},
  G.~P., {Gendreau}, K., {Ghisellini}, G., {Greiner}, J., {Hill}, J.~E.,
  {Hunsberger}, S.~D., {Krimm}, H.~A., {Kulkarni}, S.~R., {Kumar}, P.,
  {Lebrun}, F., {Lloyd-Ronning}, N.~M., {Markwardt}, C.~B., {Mattson}, B.~J.,
  {Mushotzky}, R.~F., {Norris}, J.~P., {Osborne}, J., {Paczynski}, B.,
  {Palmer}, D.~M., {Park}, H.-S., {Parsons}, A.~M., {Paul}, J., {Rees}, M.~J.,
  {Reynolds}, C.~S., {Rhoads}, J.~E., {Sasseen}, T.~P., {Schaefer}, B.~E.,
  {Short}, A.~T., {Smale}, A.~P., {Smith}, I.~A., {Stella}, L., {Tagliaferri},
  G., {Takahashi}, T., {Tashiro}, M., {Townsley}, L.~K., {Tueller}, J.,
  {Turner}, M.~J.~L., {Vietri}, M., {Voges}, W., {Ward}, M.~J., {Willingale},
  R., {Zerbi}, F.~M., and {Zhang}, W.~W., ``{The Swift Gamma-Ray Burst
  Mission},'' {\em \apj}~{\bf 611},  1005--1020 (Aug. 2004).

\bibitem{sch07}
{Schanne}, S., {Cordier}, B., {Gotz}, D., {Gros}, A., {Kestener},
P., {Le
  Provost}, H., {L'Huillier}, B., and {Mur}, M., ``{The trigger function of the
  space borne gamma-ray burst telescope ECLAIRs},'' {\em ArXiv e-prints}~{\bf
  711} (Nov. 2007).

\bibitem{cor08}
{Cordier}, B., {Desclaux}, F., {Foliard}, J., and {Schanne}, S.,
``{SVOM
  pointing strategy: how to optimize the redshift measurements?},'' in [{\em
  American Institute of Physics Conference
  Series}{\nolinebreak\hspace{0.1em}]},  {\em American Institute of Physics
  Conference Series} {\bf 1000},  585--588 (May 2008).

\bibitem{sch06}
{Schanne}, S., {Atteia}, J.-L., {Barret}, D., {Basa}, S., {Boer},
M., {Casse},
  F., {Cordier}, B., {Daigne}, F., {Klotz}, A., {Limousin}, O., {Manchanda},
  R., {Mandrou}, P., {Mereghetti}, S., {Mochkovitch}, R., {Paltani}, S.,
  {Paul}, J., {Petitjean}, P., {Pons}, R., {Ricker}, G., and {Skinner}, G.,
  ``{The ECLAIRs micro-satellite mission for gamma-ray burst multi-wavelength
  observations},'' {\em Nuclear Instruments and Methods in Physics Research
  A}~{\bf 567},  327--332 (Nov. 2006).

\bibitem{nar08}
{Nardini}, M., {Ghisellini}, G., and {Ghirlanda}, G., ``{Optical
afterglow
  luminosities in the Swift epoch: confirming clustering and bimodality},''
  {\em \mnras}~{\bf 386},  L87--L91 (May 2008).

\bibitem{ama02}
{Amati}, L., {Frontera}, F., {Tavani}, M., {in't Zand}, J.~J.~M.,
{Antonelli},
  A., {Costa}, E., {Feroci}, M., {Guidorzi}, C., {Heise}, J., {Masetti}, N.,
  {Montanari}, E., {Nicastro}, L., {Palazzi}, E., {Pian}, E., {Piro}, L., and
  {Soffitta}, P., ``{Intrinsic spectra and energetics of BeppoSAX Gamma-Ray
  Bursts with known redshifts},'' {\em \aap}~{\bf 390},  81--89 (July 2002).

\bibitem{sak08}
{Sakamoto}, T., {Hullinger}, D., {Sato}, G., {Yamazaki}, R.,
{Barbier}, L.,
  {Barthelmy}, S.~D., {Cummings}, J.~R., {Fenimore}, E.~E., {Gehrels}, N.,
  {Krimm}, H.~A., {Lamb}, D.~Q., {Markwardt}, C.~B., {Osborne}, J.~P.,
  {Palmer}, D.~M., {Parsons}, A.~M., {Stamatikos}, M., and {Tueller}, J.,
  ``{Global Properties of X-Ray Flashes and X-Ray-Rich Gamma-Ray Bursts
  Observed by Swift},'' {\em \apj}~{\bf 679},  570--586 (May 2008).

\bibitem{gru07}
{Grupe}, D., {Nousek}, J.~A., {vanden Berk}, D.~E., {Roming},
P.~W.~A.,
  {Burrows}, D.~N., {Godet}, O., {Osborne}, J., and {Gehrels}, N., ``{Redshift
  Filtering by Swift Apparent X-Ray Column Density},'' {\em \aj}~{\bf 133},
  2216--2221 (May 2007).

\bibitem{ghi07}
{Ghisellini}, G., {Ghirlanda}, G., {Nava}, L., and {Firmani}, C.,
``{``Late
  Prompt'' Emission in Gamma-Ray Bursts?},'' {\em \apjl}~{\bf 658},  L75--L78
  (Apr. 2007).

\bibitem{ghir07}
{Ghirlanda}, G., {Nava}, L., {Ghisellini}, G., and {Firmani}, C.,
``{Confirming
  the {$\gamma$}-ray burst spectral-energy correlations in the era of multiple
  time breaks},'' {\em \aap}~{\bf 466},  127--136 (Apr. 2007).

\bibitem{obr06}
{O'Brien}, P.~T., {Willingale}, R., {Osborne}, J.~P., and {Goad},
M.~R.,
  ``{Early multi-wavelength emission from gamma-ray bursts: from gamma-ray to
  x-ray},'' {\em New Journal of Physics}~{\bf 8},  121--+ (July 2006).

\bibitem{laz08}
{Lazzati}, D., {Perna}, R., and {Begelman}, M.~C., ``{X-ray
flares, neutrino
  cooled disks, and the dynamics of late accretion in GRB engines},'' {\em
  ArXiv e-prints}~{\bf 805} (May 2008).

\bibitem{tom07}
{Toma}, K., {Ioka}, K., {Sakamoto}, T., and {Nakamura}, T.,
``{Low-Luminosity
  GRB 060218: A Collapsar Jet from a Neutron Star, Leaving a Magnetar as a
  Remnant?},'' {\em \apj}~{\bf 659},  1420--1430 (Apr. 2007).

\bibitem{cam06}
{Campana}, S., {Mangano}, V., {Blustin}, A.~J., {Brown}, P.,
{Burrows}, D.~N.,
  {Chincarini}, G., {Cummings}, J.~R., {Cusumano}, G., {Della Valle}, M.,
  {Malesani}, D., {M{\'e}sz{\'a}ros}, P., {Nousek}, J.~A., {Page}, M.,
  {Sakamoto}, T., {Waxman}, E., {Zhang}, B., {Dai}, Z.~G., {Gehrels}, N.,
  {Immler}, S., {Marshall}, F.~E., {Mason}, K.~O., {Moretti}, A., {O'Brien},
  P.~T., {Osborne}, J.~P., {Page}, K.~L., {Romano}, P., {Roming}, P.~W.~A.,
  {Tagliaferri}, G., {Cominsky}, L.~R., {Giommi}, P., {Godet}, O., {Kennea},
  J.~A., {Krimm}, H., {Angelini}, L., {Barthelmy}, S.~D., {Boyd}, P.~T.,
  {Palmer}, D.~M., {Wells}, A.~A., and {White}, N.~E., ``{The association of
  GRB 060218 with a supernova and the evolution of the shock wave},'' {\em
  \nat}~{\bf 442},  1008--1010 (Aug. 2006).

\bibitem{cam08}
{Campana}, S., {Panagia}, N., {Lazzati}, D., {Beardmore}, A.~P.,
{Cusumano},
  G., {Godet}, O., {Chincarini}, G., {Covino}, S., {Della Valle}, M.,
  {Guidorzi}, C., {Malesani}, D., {Moretti}, A., {Perna}, R., {Romano}, P., and
  {Tagliaferri}, G., ``{Outliers from the mainstream: how a massive star can
  produce a gamma-ray burst},'' {\em ArXiv e-prints}~{\bf 805} (May 2008).

\bibitem{che00}
{Chevalier}, R.~A. and {Li}, Z.-Y., ``{Wind Interaction Models for
Gamma-Ray
  Burst Afterglows: The Case for Two Types of Progenitors},'' {\em \apj}~{\bf
  536},  195--212 (June 2000).

\bibitem{ghi08}
{Ghisellini}, G., {Ghirlanda}, G., {Tavecchio}, F., {Fraternali},
F., and
  {Pareschi}, G., ``{Ultra-High Energy Cosmic Rays, Spiral galaxies and
  Magnetars},'' {\em ArXiv e-prints}~{\bf 806} (June 2008).

\bibitem{tag05}
{Tagliaferri}, G., {Antonelli}, L.~A., {Chincarini}, G.,
{Fern{\'a}ndez-Soto},
  A., {Malesani}, D., {Della Valle}, M., {D'Avanzo}, P., {Grazian}, A.,
  {Testa}, V., {Campana}, S., {Covino}, S., {Fiore}, F., {Stella}, L.,
  {Castro-Tirado}, A.~J., {Gorosabel}, J., {Burrows}, D.~N., {Capalbi}, M.,
  {Cusumano}, G., {Conciatore}, M.~L., {D'Elia}, V., {Filliatre}, P.,
  {Fugazza}, D., {Gehrels}, N., {Goldoni}, P., {Guetta}, D., {Guziy}, S.,
  {Held}, E.~V., {Hurley}, K., {Israel}, G.~L., {Jel{\'{\i}}nek}, M.,
  {Lazzati}, D., {L{\'o}pez-Echarri}, A., {Melandri}, A., {Mirabel}, I.~F.,
  {Moles}, M., {Moretti}, A., {Mason}, K.~O., {Nousek}, J., {Osborne}, J.,
  {Pellizza}, L.~J., {Perna}, R., {Piranomonte}, S., {Piro}, L., {de Ugarte
  Postigo}, A., and {Romano}, P., ``{GRB 050904 at redshift 6.3: observations
  of the oldest cosmic explosion after the Big Bang},'' {\em \aap}~{\bf 443},
  L1--L5 (Nov. 2005).

\bibitem{sch00}
{Schwope}, A., {Hasinger}, G., {Lehmann}, I., {Schwarz}, R.,
{Brunner}, H.,
  {Neizvestny}, S., {Ugryumov}, A., {Balega}, Y., {Tr{\"u}mper}, J., and
  {Voges}, W., ``{The ROSAT Bright Survey: II. Catalogue of all high-galactic
  latitude RASS sources},'' {\em Astronomische Nachrichten}~{\bf 321},  1--52
  (2000).

\bibitem{gud04}
{G{\"u}del}, M., ``{X-ray astronomy of stellar coronae},'' {\em
\aapr}~{\bf
  12},  71--237 (Sept. 2004).

\bibitem{kuu06}
{Kuulkers}, E., {Norton}, A., {Schwope}, A., and {Warner}, B.,
[{\em {X-rays
  from cataclysmic variables}}{\nolinebreak\hspace{0.1em}]},  421--460, Compact
  stellar X-ray sources (Apr. 2006).

\bibitem{dow06}
{Downes}, R.~A., {Webbink}, R.~F., {Shara}, M.~M., {Ritter}, H.,
{Kolb}, U.,
  and {Duerbeck}, H.~W., ``{Catalog of Cataclysmic Variables (Downes+
  2001-2006)},'' {\em VizieR Online Data Catalog}~{\bf 5123},  0--+ (Jan.
  2006).

\end{thebibliography}
%\bibliographystyle{spiebib}   %>>>> makes bibtex use spiebib.bst

\end{document}